# Topological Dirac Semimetal $Na_3Bi$ Films in the Ultrathin Limit via Alternating Layer Molecular Beam Epitaxy

Igor V. Pinchuk[1], Thaddeus J. Asel[1], Andrew Franson[1], Tiancong Zhu[1], Yuan-Ming Lu[1], Leonard J. Brillson[1,2], Ezekiel Johnston-Halperin[1], Jay A. Gupta[1], and Roland K. Kawakami[1]

[1]*Department of Physics, The Ohio State University, Columbus OH 43210 USA*
[2]*Department of Electrical and Computer Engineering, The Ohio State University, Columbus OH 43210 USA*

## Abstract

Ultrathin films of $Na_3Bi$ on insulating substrates are desired for opening a bulk band gap and generating the quantum spin Hall effect from a topological Dirac semimetal, though continuous films in the few nanometer regime have been difficult to realize. Here, we utilize alternating layer molecular beam epitaxy (MBE) to achieve uniform and continuous single crystal films of $Na_3Bi(0001)$ on insulating $Al_2O_3(0001)$ substrates and demonstrate electrical transport on films with 3.8 nm thickness (4 unit cells). The high material quality is confirmed through *in situ* reflection high-energy electron diffraction (RHEED), scanning tunneling microscopy (STM), x-ray diffraction (XRD), and x-ray photoelectron spectroscopy (XPS). In addition, these films are employed as seed layers for subsequent growth by codeposition, leading to atomic layer-by-layer growth as indicated by RHEED intensity oscillations. These material advances facilitate the pursuit of quantum phenomena in thin films of Dirac semimetals.

Topological Dirac semimetals (TDS) are characterized by band touching and linear energy-momentum dispersion at specific locations inside the Brillouin zone, known as Dirac points.[1,2] Graphene is perhaps the most famous example of this class of materials, a 2D TDS with Dirac points present at high symmetry K and K' points.[3] Recently, materials such as $Na_3Bi$ and $Cd_2As_3$ have been demonstrated to be 3D TDS with two Dirac points situated close to the Brillouin zone center and possessing linear dispersions in all three momentum directions (Fig. 1a).[4-13] Theory predicts that quantum confinement of a 3D TDS will create a tunable bulk band gap and alternate the system between trivial insulator and quantum spin Hall insulator as a function of film thickness.[14,15] For $Na_3Bi$, this tunability requires high-quality epitaxial growth of (0001)-oriented films on insulating substrates in the ultrathin limit (few nanometers).

To date, (0001)-oriented thin films of $Na_3Bi$ have been grown on Si(111), $Al_2O_3$(0001), graphene, and $SiO_2$/Si(001) by molecular beam epitaxy (MBE),[16-22] and their band structures have been characterized by angle-resolved photoemission spectroscopy (ARPES). In addition, Fuhrer and co-workers have performed transport measurements *in situ* within their MBE chamber to avoid oxidation.[19-22] They observed mobilities as high as 7,000 $cm^2$/Vs, obtained as-grown carrier densities in the $10^{17}$ $cm^{-3}$ range, and performed gate-tunable transport with resistance maxima at the Dirac point. All of the MBE growths have been performed by codeposition in the adsorption-limited regime where excess Na flux re-evaporates and the growth rate is determined by the Bi flux. However, while transport studies have been performed on thicker $Na_3Bi$ films on insulating substrates (20 nm or greater[19-22]), transport measurements in the ultrathin regime have not been reported.

In this work, we report the growth of high quality, continuous $Na_3Bi$ films on insulating $Al_2O_3$(0001) in the ultrathin regime by employing alternating layer MBE (AL-MBE) and demonstrate ohmic conductivity in films with 3.8 nm thickness (4 unit cells). We find that sequential growth of atomic layers of Na and Bi at room temperature results in (0001)-oriented $Na_3Bi$ films with uniform surface coverage at very low thickness. Reflection high-energy electron diffraction (RHEED) and x-ray photoelectron spectroscopy (XPS) measurements reveal that single crystal films are formed after just 2 unit cells of deposition. Scanning tunneling microscopy (STM) shows smooth films with continuous coverage, and van der Pauw measurements demonstrate charge transport on 3.8 nm films. X-ray diffraction (XRD) measurements on a 8 unit cell film exhibit a peak in the θ-2θ scan corresponding to $Na_3Bi$(0002). Furthermore, these AL-MBE films can also serve as seed layers for depositing high quality thick films using codeposition, as revealed by STM and the observation of RHEED oscillations indicating atomic layer-by-layer growth. This new growth method will enable investigation of quantum confinement and other novel physical phenomena in the ultrathin limit of Na$_3$Bi thin films.

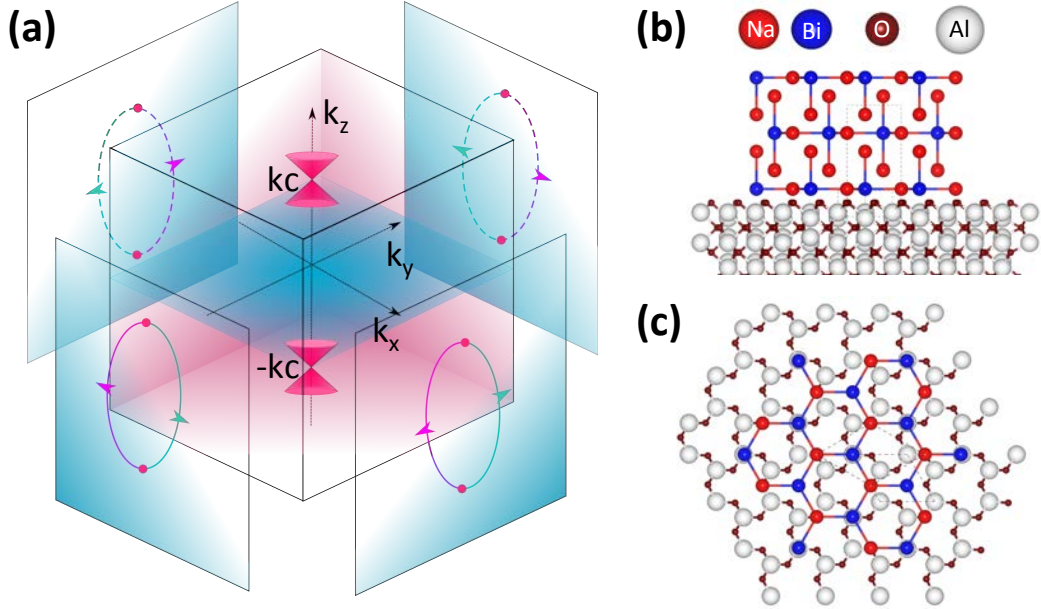

**FIG. 1.** (a) Momentum-space diagram of Na$_3$Bi, showing two Dirac points ±kc along the k$_z$ axis. This dispersion leads to non-trivial Fermi arc surface states on the [0100] surfaces (pink and blue arcs). (b) Side view Na$_3$Bi/Al$_2$O$_3$ bilayer along Na$_3$Bi [1000] azimuth. (c) Top view along Na$_3$Bi [0001] azimuth showing the alignment of honeycomb NaBi layer over the Al$_2$O$_3$(0001) surface. Note the good lattice match due to the 30° relative rotation between the unit cells.

Na$_3$Bi ultrathin films are grown on Al$_2$O$_3$(0001) substrates in an MBE chamber with a base pressure of $8\times10^{-11}$ Torr. The Al$_2$O$_3$(0001) substrates (MTI crystal) are prepared by cleaning with isopropyl alcohol and annealing in air at 1000 °C for 3 hours to saturate oxygen vacancies and smooth the surface. The substrates are then annealed in vacuum at 600 °C for 30 min. Elemental Na (99.95%, Alfa Aesar) and Bi (99.999%, Alfa Aesar) are evaporated from thermal effusion sources whose growth rates are calibrated by a quartz deposition monitor. For codeposition in the adsorption-limited regime, typical deposition rates are ~2.5 nm/min for Na and ~0.17 nm/min for Bi, where the excess Na is re-evaporated due to the elevated substrate temperature (120 - 170 °C). For AL-MBE growth, the substrate is at room temperature and the Na and Bi have typical growth rates of ~0.2 nm/min. The sample temperature is monitored by a transferrable thermocouple that directly measures the temperature of the tantalum sample paddle. Since Na$_3$Bi will oxidize when exposed to air, most characterizations are performed in air-free environments. For *in situ* characterization by XPS and STM, samples are transferred from the MBE chamber into the respective systems using an ultrahigh vacuum (UHV) suitcase;

for transport characterization, samples are transferred from UHV to a glove box using a custom-made UHV-compatible sample vessel that is carried into the glove box through its antechamber; for XRD characterization, samples are transferred from UHV to the glove box and loaded into an air-free XRD sample holder to avoid air exposure during the measurement.

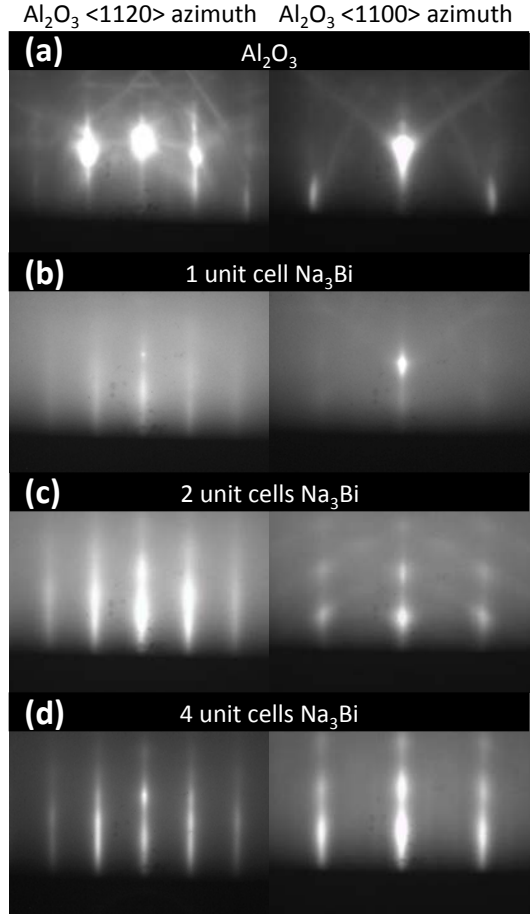

**FIG. 2.** (a) RHEED images of $Al_2O_3$ surface prior to $Na_3Bi$ growth along two pricipal azimuth directions. (b – d) RHEED images after subsequent deposition of 1, 2 and 4 unit cells of $Na_3Bi$ using AL-MBE growth at room temperature. Diffraction patterns remain largely unchanged for higher thicknesses.

$Na_3Bi$ has a hexagonal crystal structure (P63/mmc space group) with lattice constants $a$ = 0.5448 nm and $c$ = 0.9655 nm.[23] As shown in Figure 1b, it consists of honeycomb NaBi layers which are sandwiched between hexagonal Na layers above and below along the $c$-axis, making up half of a unit cell. Since the underlying $Al_2O_3$(0001) substrate has an in-plane lattice constant of $a$ = 0.4785 nm, $Na_3Bi$ films with (0001) orientation and 30° in-plane rotation have a lattice mismatch of only 1.4% (Figure 1c).

The AL-MBE growth of $Na_3Bi$ on $Al_2O_3$(0001) is performed at room temperature and consists of alternating the growth of a flux-matched NaBi monolayer followed by pure Na bilayer and repeating this sequence. Figure 2 summarizes the evolution of RHEED patterns throughout the AL-MBE growth of $Na_3Bi$. Figure 2a shows the $Al_2O_3$(0001) substrate prior to growth along the <1120> and <1100> azimuths, indicating a clean and smooth starting surface. After depositing 1 unit cell of $Na_3Bi$, a faint RHEED pattern of $Na_3Bi$(0001) appears along both directions while the $Al_2O_3$ Kikuchi lines are suppressed (Figure 2b). After depositing an additional unit cell, the diffraction pattern is very clear along both directions with no hint of the underlying substrate (Figure 2c). Terminating the AL-MBE growth after 4 unit cells leads to the patterns shown in Figure 2d, which remain largely unchanged even at higher thickness.

To characterize the film morphology, we perform a combination of *in situ* STM and *ex situ* AFM measurements. Figure 3a shows an AFM image of an $Al_2O_3$(0001) substrate after a 1000 $^{o}$C air anneal which produces flat surfaces with ordered atomic steps. Following the growth of 5 unit cells of $Na_3Bi$ by AL-MBE, we image the topography by *in situ* STM (Figure 3b) under constant current conditions (200 pA tip current, 200 mV sample bias). The images exhibit a smooth surface morphology, with terraces having typical widths of ~5 nm and step heights of ~0.5 nm corresponding to a single monolayer (half unit cell). Moreover, the films are macroscopically flat, with height variation of ~1 nm (Fig. 3b inset). To compare the AL-MBE growth with the conventional codeposition growth,[19-22] we perform codeposition growth directly onto an $Al_2O_3$(0001) substrate at 120 $^{o}$C (the Na:Bi flux ratio is ~20:1). Our attempts to perform *in situ* STM failed due to tip crash, which suggests that the deposited films are not continuous and thus not electrically conducting. Therefore, we utilize *ex situ* AFM measurements to perform the comparison. Figures 3c and 3d show AFM images of 5 unit cell $Na_3Bi$ films deposited by AL-MBE and codeposition growth, respectively. It is worthwhile to note that the air exposure does cause the AL-MBE film to be rougher in the AFM image (Figure 3c) compared to the *in situ* STM image (Figure 3b), as one would expect from sample oxidation. Nevertheless, the AFM comparison provides a clear indication of the improved morphology and continuous film structure for AL-MBE growth (Figure 3c) as compared to the codeposition growth which is highly non-uniform with bare substrate patches separating disconnected islands (Figure 3d).

The continuity of the AL-MBE films is further verified by electrical transport measurements which are performed inside a glove box. The room temperature resistivity measurement of a 4 unit cell $Na_3Bi$ film shows a linear I-V curve with a resistivity of $\sim 7 \times 10^{-4}$ Ω m (Figure 3e). This is higher than $\sim 5 \times 10^{-5}$ Ω m previously reported for thicker 20 nm films, which may suggest band gap opening or the importance of surface scattering in the ultrathin films. Further studies are needed to understand the sources of electron momentum scattering in ultrathin films.

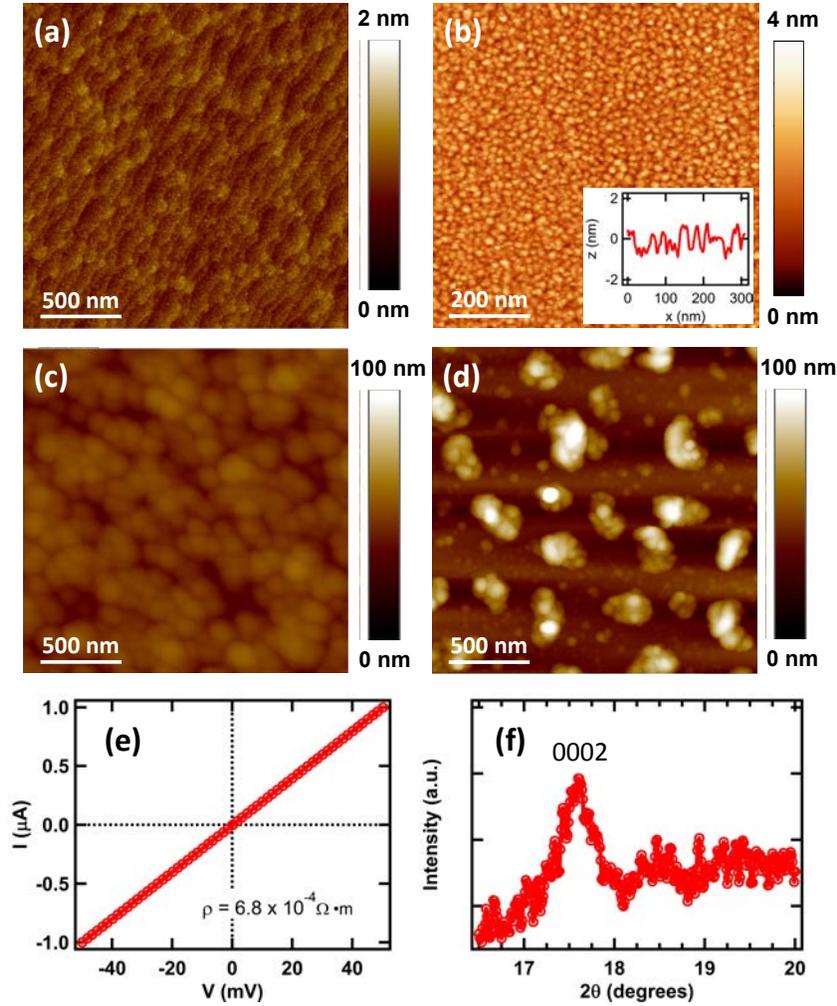

**FIG. 3.** (a) A 2 μm x 2 μm AFM scan of annealed $Al_2O_3$ substrate before $Na_3Bi$ deposition, showing atomic steps and a very flat surface. (b) STM scan of 5 unit cell AL-MBE $Na_3Bi$ taken immediately after growth. Inset: a cross sectional line cut showing height variations of ~1 nm. (c) AFM scan of 5 unit cell AL-MBE $Na_3Bi$. (d) AFM scan of 5 unit cell $Na_3Bi$ grown at 120 °C using codeposition. Note that scans in (c) and (d) were taken after the $Na_3Bi$ surface was exposed to air and thus oxidized. (e) A linear van der Pauw I-V curve taken at 300 K giving an average resistivity of $6.8\times10^{-4}$ Ω m. (f) θ-2θ x-ray diffraction scan of AL-MBE $Na_3Bi$ (8 unit cell)/$Al_2O_3$ substrate heterostructure, showing a clear $Na_3Bi$ 0002 peak.

Additional structural characterization is provided by XRD performed on a 8 unit cell AL-MBE film (Figure 3f). The θ-2θ scan shows a peak at 2θ = 17.5° which corresponds to the (0002) diffraction peak of $Na_3Bi$ and thus verifies the (0001) crystallographic orientation of the film.

*In situ* XPS characterization provides insight into the chemical composition of the $Na_3Bi$ films (Figure 4). Survey scans from 4 unit cell films grown by AL-MBE (Figure 4a, green dashed line) and codeposition (Figure 4a, red solid line) show peaks from Na 1s, Bi 4f, O 1s, and Al 2p core levels, as well as Auger peaks (Na KLL). The Al 2p and O 1s peaks are expected due to the finite escape depth of photoelectrons from the $Al_2O_3$ substrate.[24] Analysis of the Na 1s peaks (Figure 4b) and Bi 4f peaks (Figure 4c) from the two samples reveal a stoichiometry of $Na_{3.76}Bi$ for the AL-MBE growth and $Na_{4.54}Bi$ for the codeposition, indicating that both films are Na rich. In comparing the two samples, the AL-MBE growth has sharper and taller Na and Bi peaks, while the codeposition growth has relatively smaller Na and Bi peaks. The spectra on codeposited films also have shoulder peaks at 1074.20 eV, 164.24 eV, and 158.72 eV, which are indicative of $Na_2O$ and $Bi_2O_3$ formation.[25,26] Furthermore, comparing the relative heights of the O 1s peaks (Figure 4d) and the Al 2p peaks (Figure 4e) provides additional evidence for smoother film morphology by AL-MBE growth. The much stronger O and Al peaks for codeposition compared to AL-MBE growth indicates an increased contribution from the $Al_2O_3$ substrate due to patchy growth for codeposition, which is consistent with the discontinuous island morphology observed in the AFM image (Figure 3d). These results indicate that AL-MBE results in material of higher quality than codeposition in the ultrathin film regime.

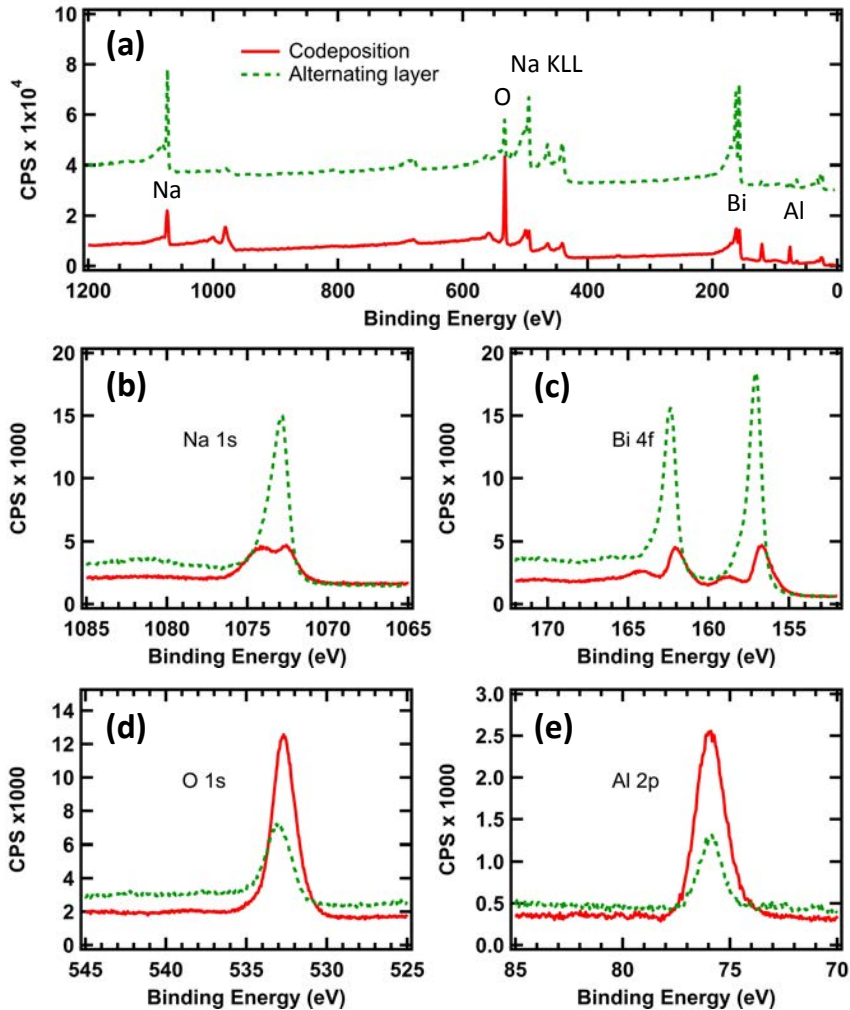

**FIG. 4.** (a) XPS survey scan of 4 unit cell $Na_3Bi$ films grown using room temperature AL-MBE (green dashed line) and 120 °C codeposition (red solid line). The AL-MBE scan is vertically offset for clarity. The Al 2p peak from $Al_2O_3$ is used for global calibration of binding energies. (b, c) XPS region scans of Na 1s and Bi 4f core levels, respectively. Note the lack of partial oxidation shoulders in the AL-MBE sample. (d, e) XPS region scans of O 1s and Al 2p core levels, respectively. Stronger Al and O signals combined with weaker Na and Bi signals suggest a more uniform coverage for AL-MBE growth.

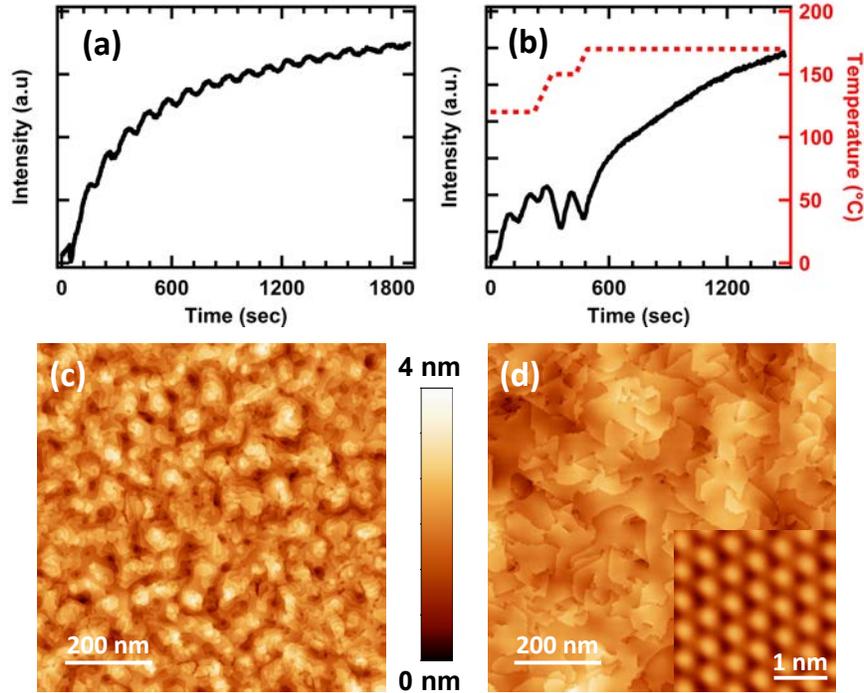

**FIG. 5.** (a) RHEED intensity oscillation as a function of time during 120 °C codeposition of $Na_3Bi$ on top of a 4 unit cell AL-MBE seed layer. (b) RHEED intensity oscillation (left axis) taken during codeposition of $Na_3Bi$ on top of a 4 unit cell AL-MBE seed layer as the temperature is ramped from 120 °C to 170 °C (right axis). Note that the oscillations die out between 150 °C and 170 °C. (c, d) STM scans of 35 nm $Na_3Bi$ deposited at 120 °C and 170 °C, respectively. Both films were deposited on top of 5 unit cell AL-MBE seed layers. Inset: Atomic resolution scan for the 170 °C sample.

In addition to ultrathin films, the improved morphology of AL-MBE enables the synthesis of uniform thicker films by using the AL-MBE film as a seed layer for subsequent codeposition growth. The advantage of codeposition in the adsorption-limited regime is the better control over stoichiometry by re-evaporation of excess Na. We have investigated this growth mode at two different temperatures, 120 °C and 170 °C. In both cases, we initiate growth on $Al_2O_3$(0001) with a 4 unit cell $Na_3Bi$ film grown by AL-MBE at room temperature. The substrate temperature is then ramped to 120 °C, and codeposition growth is initiated by opening the Na and Bi shutters (the Na:Bi flux ratio is ~20:1). Notably, we observe RHEED intensity oscillations for growth at 120 °C (Fig. 5a) which indicates atomic layer-by-layer growth. Interestingly the oscillation period is a full unit cell, or two monolayers, which indicates that the lattice structure of the unit cell remains intact. For the 170 °C growth, the substrate temperature is ramped from 120 °C to 170 °C during the initial stages of growth, as shown in Figure 5b. We observe RHEED oscillations during the initial stages of growth, but they disappear when the substrate temperature exceeds 150 °C. STM measurements compare the morphology of 35 nm thick films grown under these two conditions. Figure 5c shows a typical STM image for a 120 °C sample, showing ~0.5 nm atomic steps and typical terrace widths of 40 nm. Figure 5d shows a typical STM image for a 170 °C sample, showing ~0.5 nm atomic steps and typical terrace widths of 150 nm. The larger terraces for the 170 °C suggests that disappearance of RHEED oscillations is due to increased mobility of the atoms and possible step-flow growth at the higher temperature, and not an indicator of poor film quality. The inset of Figure 5d is a high resolution image of the atomic lattice whose measured in-plane lattice constant is 0.55 nm, which is consistent with $Na_3Bi$ (0.5448 nm). Finally, XPS spectra taken on films grown at 170 °C indicate a stoichiometry of $Na_{3.50}Bi$, which is slightly better than the AL-MBE grown films in Figure 4. These characterizations indicate the high quality and uniformity of thick $Na_3Bi$ films synthesized by codeposition on top of AL-MBE seed layers.

In summary, we demonstrate that alternating layer MBE growth at room temperature can be used to synthesize uniform and continuous $Na_3Bi$ films on $Al_2O_3$(0001) substrates in the ultrathin regime (few nm). The morphology is greatly improved compared to codeposition of $Na_3Bi$ films, which grow as disconnected islands when deposited directly on $Al_2O_3$(0001). We demonstrate in-plane electrical conductance on films as thin as 3.8 nm (4 unit cells), which enables transport studies of topological and quantum phenomena in these systems. In addition, the ultrathin AL-MBE films can be used as seed layers for atomic layer-by-layer growth by codeposition in the adsorption-limited regime. These results facilitate the investigation of $Na_3Bi$ films in the ultrathin limit, where the effects of quantum confinement can lead to novel topological phenomena such as thickness-tunable quantum spin Hall effect.

**Acknowledgements**

We acknowledge the technical assistance of Guanzhong Wu. Funding for this research was provided by the Center for Emergent Materials: an NSF MRSEC under award number DMR-1420451.